\documentclass[aps,prl,amsmath,amsfonts,amssymb,notitlepage,twocolumn,superscriptaddress,footinbib,floatfix]{revtex4-2}
\usepackage{CJK}
\usepackage{bm}
\usepackage{soul}
\usepackage[caption=false,subrefformat=parens,labelformat=parens]{subfig}
\usepackage{color} 
\usepackage[table,dvipsnames]{xcolor}
\usepackage[colorlinks=true,linkcolor=black]{hyperref}
\usepackage{enumitem}
\usepackage{ulem}
\usepackage{xspace}
\usepackage{tikz}
\usepackage{soul}

\definecolor{col1}{rgb}{0, 0.6, 0}
\definecolor{col2}{rgb}{0.9, 0.7, 0.3}
\definecolor{col3}{rgb}{0.2,0.3,0.9}
\definecolor{col4}{rgb}{1,1,0.7}
\definecolor{col5}{rgb}{0.8,1,0.7}
\definecolor{col6}{rgb}{0.8,0.9,1}

\usepackage{graphicx}
\graphicspath{{FIG/}}
\usepackage{multirow}
\usepackage{braket} 
\usepackage{booktabs} 

\AtBeginDocument{
\heavyrulewidth=.08em
\lightrulewidth=.05em
\cmidrulewidth=.03em
\belowrulesep=.65ex
\belowbottomsep=0pt
\aboverulesep=.4ex
\abovetopsep=0pt
\cmidrulesep=\doublerulesep
\cmidrulekern=.5em
\defaultaddspace=.5em
}

\newcommand{\prlsection}[1]{\textbf{{#1}}}

\makeindex

\begin{document} 
\begin{CJK*}{UTF8}{gbsn} 
\title{
Scattering on the Worldvolume:
\\
Amplitude Relations in Brower-Goddard String Models
}
 \author{N. Emil J. Bjerrum-Bohr}  
\affiliation{Niels Bohr International Academy, Niels Bohr Institute, Blegdamsvej 17, DK-2100 Copenhagen  {{\O}}, Denmark}
 \author{Christian Baadsgaard Jepsen}
 \affiliation{School of Physics, Korea Institute for Advanced Study, Seoul 02455, Korea}
\begin{abstract}
We investigate the Brower-Goddard extension of the Veneziano and Virasoro-Shapiro four-point amplitudes obtained by generalizing the Koba-Nielsen integrals to $d$-dimensional conformally invariant integrals. The amplitudes derived from this framework exhibit polynomial residues and can be shown to adhere to polynomial bounds at high energies. In odd dimensions, the amplitudes decompose into sums of three partial amplitudes, enabling the formulation of general amplitude relations that subsume the Kawai-Lewellen-Tye (KLT) formula as a particular case. The amplitudes contain multiple tachyons in their spectra. Still, we demonstrate that their residues comply with the positivity conditions mandated by unitarity for spacetime dimensions at or below critical values $D_\text{crit}(d)$, where $D_\text{crit}(6)=26$ and $D_\text{crit}(\infty)=10$. In closing, we contemplate physical applications for membranes and potential extensions of the formalism.
\end{abstract}
\maketitle
\end{CJK*}

\prlsection{Introduction} 
New physical insights sometimes come about through analysis of scattering phenomena from the ground up by examining the analytic properties of amplitudes. In this manner, the Veneziano amplitude \cite{Veneziano:1968yb} played a pivotal role in the advent of string theory. In this presentation, we examine the properties of the four-point amplitudes derived by expanding the Koba-Nielsen integral measure in string theory \cite{Koba:1969rw} to higher dimensions,
\begin{align}
\label{intgeneral}
A^{(d)}(s,t)= \int  d\Omega_4^{(d)}\,\prod_{i=1}^{3}\prod_{j=i+1}^4|\vec{x}_j-\vec{x}_i|^{2k_i\cdot k_j} \,,
\end{align}
where the generalized integration measure is provided by
\begin{align}
\int d\Omega_4^{(d)} \equiv\,& 
|\vec{x}_a-\vec{x}^{\,0}_b|\,
|\vec{x}_b-\vec{x}^{\,0}_c|\,
|\vec{x}_c-\vec{x}^{\,0}_a|
\,
\bigg[\prod_{i=1}^4\int_{\mathbb{R}^d} d^d\vec{x}_i\bigg]
\nonumber\\
&
\delta(\vec{x}_a-\vec{x}_a^{\,0})\,
\delta(\vec{x}_b-\vec{x}_b^{\,0})\,
\delta(\vec{x}_c-\vec{x}_c^{\,0})
\,.
\end{align}
Here $\vec{x}_a$, $\vec{x}_b$, and $\vec{x}_c$ are any three of the four vector-valued integration variables $\vec{x}_1$ to $\vec{x}_4$, and $\vec{x}^{\,0}_a$, $\vec{x}^{\,0}_b$, and $\vec{x}^{\,0}_c$ are any three distinct vectors in $\mathbb{R}^d$. We denote the dimensionless momenta with $\alpha'$ absorbed into them by $k_i$. They are conserved and satisfy a tachyonic on-shell condition, which reads $(k_i)^2 = d$ in a mostly-positive signature. This mass condition ensures that the $PSL(2,\mathbb{R})$ and $PSL(2,\mathbb{C})$ symmetry for $d\in\{1,2\}$ is replaced with $d$-dimensional conformal symmetry for the dimensionally uplifted amplitudes. By manipulating known gamma function integral identities assuming analytic continuation, as outlined in \hyperref[appendix:A]{Appendix A}, the integral in \eqref{intgeneral} can be readily performed, resulting in the identification  
\begin{align}
\label{Ad}
A^{(d)}(s,t)=\pi^{d/2}\frac{\Gamma(\frac{-d-s}{2})\Gamma(\frac{-d-t}{2})\Gamma(\frac{-d-u}{2})}{\Gamma(\frac{2d+s}{2})\Gamma(\frac{2d+t}{2})\Gamma(\frac{2d+u}{2})}\displaystyle\,,
\end{align}
where the Mandelstam invariants are defined by $s\equiv -(k_1+k_2)^2$, $t\equiv -(k_1+k_3)^2$ with $s+t+u=-4d$.  \hyperref[appendix:B]{Appendix B} presents an equivalent expression for $A^{(d)}(s,t)$ in terms of a type of zeta function. For $d=1$, one recovers, as expected, the Veneziano amplitude,
\begin{align}
A^{(1)}(s,t) =\, &
B(-1-s,-1-t)+B(-1-s,-1-u)
\nonumber \\
\label{d1split}
&+B(-1-t,-1-u)\,,
\end{align}
where $B(x,y)=\frac{\Gamma(x)\Gamma(y)}{\Gamma(x+y)}$ is the Euler beta function. Setting $d=2$ yields the Virasoro-Shapiro amplitude \cite{Virasoro:1969me,Shapiro:1970gy}. 

The extended amplitudes $A^{(d)}(s,t)$, both in the conformally symmetric integral form \eqref{intgeneral} and in the explicit gamma function form \eqref{Ad}, were first written down in 1971 by Brower and Goddard \cite{brower1971generalized}, who also presented evidence for the correct factorization of the higher-point amplitudes but cautioned that the for $d\neq \{1,2\}$, the underlying theory might not be ghost-free. The formulas \eqref{intgeneral} and \eqref{Ad} were independently rediscovered in 1993 by Natsuume \cite{natsuume1993natural} after his advisor Polchinski suggested that he study the problem of generalizing bosonic worldsheet amplitudes to (hyper)worldvolumes. Ref.~\cite{natsuume1993natural} also provided early indications that $A^{(d)}(s,t)$ is consistent with unitarity for general values of $d$. However, based on observations of a natural extension of $A^{(d)}(s,t)$ to six-particle scattering, Green and Thorn \cite{green1991continuing} reported indications of ghosts in the spectrum unless $d\in\{1,2\}$, and little has been written on the subject since. (See, however, Refs.~\cite{siegel2016f,siegel2020s} for more recent work by Siegel on supersymmetrized, higher-dimensional Koba-Nielsen integrals.)

In light of the massive developments that have taken place over the past decades in the study of $S$-matrices, we here subject the amplitudes $A^{(d)}(s,t)$ with $d>2$ to renewed and closer scrutiny, discovering in the process a number of appealing features from the perspective of physics. Our initial focus will be on examining the pole structure and high energy asymptotics of the generalized amplitudes. Subsequently, we demonstrate the feasibility of splitting $A^{(d)}(s,t)$   for odd $d$ into a sum of three partial amplitudes, enabling us to derive a KLT-like formula for the generalized amplitude. Finally, we investigate the partial wave decompositions and critical dimensions of the generalized amplitudes before concluding and looking ahead. The consistency of higher-point amplitudes in string theory \cite{arkani2024multiparticle} is a crucial requirement and warrants further investigation, but we are here devoted to four-point amplitudes.
 \\
 
 \prlsection{Polynomial Residues and High Energy Limits} 
\label{sec:residues}
The $s$-channel poles of $A_p^{(d)}(s,t)$ are located at the points $s = m_n^2 \equiv -d+2n$, where $n$ is a non-negative integer. It implies the presence of $\lceil \frac{d}{2}\rceil$ tachyons in the theory's spectrum. For $d>2$, the spectrum suggests the possibility of a potential with multiple unstable directions. Focusing on the $s$-channel residues, we can express them in terms of Pochhammer symbols:
\begin{align}
\underset{s=m^2_n}{\text{Res}}A^{(d)} =
-2\pi^{\frac{d}{2}}(-1)^n
\frac{(-\frac{d+t}{2}-n)_n(\frac{t}{2}+d)_n}{n!\,\Gamma(\frac{d}{2}+n)}
\,,
\label{residues}
\end{align}
which are polynomials in $t$ as required by locality. The $t$-dependencies of the residues indicate that the states at mass level $m_n^2$ carry even spins 0 to $2n$, with tachyonic states carrying spins 0 to $2\left\lceil\frac{d}{2}\right\rceil-2$.

The amplitudes exhibit benign high-energy asymptotics in that they are polynomially bounded. Invoking Stirling's approximation, one can check that in the Regge limit of large $s$, fixed $t$, the generalized amplitudes exhibit the following asymptotics
\begin{align}
&
A^{(d)}(s,t) \sim \pi^{d/2}(s/2)^{d+t}(-1)^d
\frac{\sin(\pi\frac{s+t}{2})}{\sin(\pi\frac{s+d}{2})}
\frac{\Gamma(-\frac{t+d}{2})}{\Gamma(\frac{t}{2}+d)}\,,
\label{asymp1}
\end{align}
where the symbol ``$\sim$" indicates that the ratio tends to one in the given limit. Meanwhile, setting $t=-\frac{s+4d}{2}(1+\cos\theta)$, we find that in the limit of large $s$ and fixed $\cos\theta$, the asymptotics are given by
\begin{align}
&
\label{asymp2}
A^{(d)}\big(s,t\big) \sim 
-2
\left(\frac{\pi}{2s}\right)^{d/2}
\big[f(\theta)\big]^{s/2}g(\theta)
\\
\nonumber
&\hspace{12mm}
\frac{\sin\big(\pi\frac{s(1+\cos\theta)+4d\cos\theta}{4}\big)\sin\big(\pi\frac{s(1-\cos\theta)-4d\cos\theta}{4}\big)}{\sin\big(\pi\frac{s+d}{2}\big)}\,,
\end{align}
where we have introduced the two shorthands
\begin{align}
& f(\theta) \equiv \frac{(1-\cos\theta)^{1-\cos\theta}(1+\cos\theta)^{1+\cos\theta}}{4}\,,
\\ \nonumber
& g(\theta) \equiv (1+\cos\theta)^{d(1+4\cos\theta)/2}
(1-\cos\theta)^{d(1-4\cos\theta)/2}\,.
\end{align}
(It can readily be checked that the function $f(\theta)$ satisfies the bounds $\frac{1}{4} \leq f(\theta) \leq 1\,,$ so that the factor of $\big[f(\theta)\big]^{s/2}$ in \eqref{asymp2} signifies exponential decay.)

Finally, in the limit of large $s$ and $t$, the asymptotics of the amplitudes  are given by
\begin{align}
A^{(d)}(s,t) \sim -\frac{(2\pi)^{\frac{d}{2}}\sin(\pi\frac{2d+s+t}{2})(s+t)^{s+t+\frac{5d}{2}}}{2\sin(\pi\frac{s+d}{2})\sin(\pi\frac{t+d}{2})\,(st)^{\frac{3d}{2}}s^{s}\,t^{t}},
\end{align}
which is consistent with the main result of Ref.~\cite{caron2017strings}. This paper proves that for any amplitude $A(s,t)$ containing higher-spin particles in its spectrum, under mild technical assumptions, the leading piece at large $s$ and $t$ of $\log A(s,t)$ is given by
$\alpha'\Big((s+t)\log(s+t)-s\log s-t\log t\Big)$.
\\

\prlsection{Partial Amplitudes, Monodromy and KLT} 
Just as the full Veneziano amplitude is expressible as a sum of three partial amplitudes in \eqref{d1split}, so too, it turns out, are the amplitudes with $d$ odd. By analyzing the residues of \eqref{Ad}, one discovers that $A^{(d)}(s,t)$ can be built up out of generalized partial amplitudes given by
\begin{align}
&A_p^{(d)}(s,t)=
-(16\pi)^{\frac{d}{2}}
\frac{\Gamma(-s-2d+2)\Gamma(-t-2d+2)}{\sqrt{1024\pi}\,\Gamma(-s-t-2d-1)}
\\[-10pt] \nonumber
&\times
\big(\frac{s+d+2}{2}\big)_{\frac{d-3}{2}}
\big(\frac{t+d+2}{2}\big)_{\frac{d-3}{2}}
\big(\frac{s+t+2d+3}{2}\big)_{\frac{d-3}{2}}\,.
\end{align}
Adding up the partial amplitudes for the three scattering channels, one finds that
\begin{align}
&\hspace{16mm}
A_p^{(d)}(s,t)
+
A_p^{(d)}(s,u)
+
A_p^{(d)}(t,u)
=
\\ \nonumber
&
\begin{cases}
A^{(d)}(s,t) 
&\text{for }d\text{ odd}\,,
\\[8pt]
\Big(1+\frac{2}{1+\cos(\pi s)+\cos(\pi t)+\cos(\pi u)}\Big)A^{(d)}(s,t) 
&\text{for }d\text{ even}\,.
\end{cases}
\end{align}
While the partial amplitudes contain all the poles of the full amplitudes, they also contain additional poles that cancel when adding together the partial amplitudes. For even $d$, the partial amplitudes have bi-infinite sequences of extra poles, all with non-polynomial residues. The odd $d$ partial amplitudes $A_p^{(d)}(s,t)$ contain new poles situated at $s=-2d+2+2n$ for $n$ a non-negative integer. Of these new odd $d$ poles, those with $s> -4$ have polynomial residues, while those with $s\leq -4$ have residues that are rational in $t$ rather than polynomial. This pathology can be cured by adding an $(s\leftrightarrow t)$-symmetric correction term $R^{(d)}(s,t)$ that is a rational function in $s$ and $t$:
\begin{align}
\label{AandA}
\mathcal{A}_p^{(d)}(s,t) = A_p^{(d)}(s,t)  + R^{(d)}(s,t)\,.
\end{align}
The decomposition of the full amplitude into partial amplitudes remains valid for the corrected partial amplitudes $\mathcal{A}_p^{(d)}(s,t)$ as long as
\begin{align}
R^{(d)}(s,t)+R^{(d)}(s,u)+R^{(d)}(u,t)=0\,.
\end{align}
The above-mentioned conditions do not suffice to uniquely determine the corrections $R^{(d)}(s,t)$ unless their behaviour at infinity is known. But there does exist a physically motivated method of deriving the corrected amplitudes $\mathcal{A}_p^{(d)}(s,t)$. The idea is to decompose the integral in \eqref{intgeneral} into three fundamental domains related by conformal symmetry and to identify each integral with a partial amplitude. We describe this method in detail in \hyperref[appendix:C]{Appendix C}. The first few correction terms are given by
\begin{align}
&R^{(1)}(s,t)=0\,, \hspace{10mm} R^{(3)}(s,t)=\frac{2\pi}{(s+4)(t+4)}\,,
\\ 
&R^{(5)}(s,t)=
-\frac{\pi^2}{2(s+4)(t+4)}
+\frac{2\pi^2}{(s+6)(t+6)}
+
\\ \nonumber
&
\frac{8\pi^2}{(s\hspace{-0.3mm}+\hspace{-0.3mm}4)(t\hspace{-0.3mm}+\hspace{-0.3mm}4)(s\hspace{-0.3mm}+\hspace{-0.3mm}8)(t\hspace{-0.3mm}+\hspace{-0.3mm}8)}
\hspace{-0.5mm}-\hspace{-0.5mm}\frac{4\pi^2}{(s\hspace{-0.3mm}+\hspace{-0.3mm}6)(t\hspace{-0.3mm}+\hspace{-0.3mm}6)(s\hspace{-0.3mm}+\hspace{-0.3mm}8)(t\hspace{-0.3mm}+\hspace{-0.3mm}8)}\,.
\end{align}
It is possible to write-down general closed form expressions for the correction terms. They are given by sums over two types of terms: terms with the reciprocal linear in $s$ and in $t$, and terms with the reciprocal quadratic in $s$ and in $t$,
\begin{align}
\nonumber\\[-30pt]
\label{Rformula}
&R^{(d)}(s,t) = \sum_{M=0}^{\frac{d-3}{2}}\frac{a^{(d)}(M)}{(s+4+2M)(t+4+2M)}
+
\\[-10pt] \nonumber
&\hspace{6mm}\sum_{M=0}^{\frac{d-5}{2}}
\frac{1}{(s+2d-2-2M)(t+2d-2-2M)}
\\ \nonumber
&\times\sum_{n=0}^{d-4-2M}
\frac{b^{(d)}(M,n)}{(s+4+2M+2n)(t+4+2M+2n)}\,,
\end{align}
with the coefficients for the two types of terms given by
\begin{align}
\label{aandbformula}
&a^{(d)}(M)=
-\frac{(-\pi)^{\frac{d-1}{2}}4^{M+2}\left(\frac{3-d}{2}\right)_M}{2^{d}\Gamma(\frac{d-1}{2})M!}\,,
\\ \nonumber
&b^{(d)}(M,n)\hspace{-0.6mm}=\hspace{-0.6mm}
\frac{64(-\pi)^{\frac{d-1}{2}}\hspace{-0.4mm}(3\hspace{-0.4mm}-\hspace{-0.4mm}d)_{\hspace{-0.5mm}M}\hspace{-1mm}\left(\frac{4-d+2n}{2}\right)_{\hspace{-0.5mm}M}\hspace{-0.9mm}(3\hspace{-0.6mm}+\hspace{-0.6mm}M\hspace{-0.6mm}-\hspace{-0.6mm}d)_n}{2^{d}(d\hspace{-0.4mm}-\hspace{-0.4mm}n\hspace{-0.4mm}-\hspace{-0.4mm}3\hspace{-0.4mm}-\hspace{-0.4mm}2M)^{-2}\Gamma(\frac{d-1}{2})n! M!\left(\frac{4-d}{2}\right)_M}.
\end{align}
The formulas we present in the remainder of this section are phrased in terms of $A_p^{(d)}(s,t)$, which results in simpler equations. Using \eqref{AandA}, the formulas for $d$ odd can all be recast in terms of the more physical, corrected partial amplitudes $\mathcal{A}_p^{(d)}(s,t)$, but  the resulting equations get increasingly unwieldy as the correction term $R^{(d)}(s,t)$ grows convoluted with increasing $d$. Such is the price of working in higher dimensions.

The Veneziano partial amplitudes are related to one another via a set of monodromy relations \cite{plahte1970symmetry,bjerrum2009minimal,stieberger2009open}. Identical relations also exist for the odd $d$ partial amplitudes,
\begin{align}
A_p^{(d)}(s,u)
=
A_p^{(d)}(s,t)\times
\begin{cases}
\frac{\sin(\pi t)}{\sin(\pi u)} &\text{ for $d$ odd,}
\\[5pt]    
\frac{1+\cos(\pi u)}{1+\cos(\pi t)} &\text{ for $d$ even.}
\end{cases}
\end{align}
An important property of the Veneziano and Virasoro-Shapiro amplitudes are the KLT relations \cite{kawai1986relation,Bern:1998sv,bjerrum2011momentum}, a type of open-closed string relation, which in our notation reads
\begin{align}
A^{(2)}(2s,2t)=
-\frac{\sin(\pi s)\sin(\pi t)}{\sin\big(\pi(s+t)\big)}
A_p^{(1)}(s,t)^2\,.
\end{align}
This relation is but the first instance of an amplitude doubling formula, valid for both even and odd $d$:
\begin{align}
&A^{(2d)}(2s,2t)
=\frac{-1}{2^{d-1}}
\frac{\sin(\pi s)\sin(\pi t)}{\sin\big(\pi(s+t)\big)}
A_p^{(d)}(s,t)^2
\\ \nonumber
&\hspace{13mm}\times
\frac{
\big(\frac{s+d+1}{2}\big)_{\frac{d-1}{2}}
\big(\frac{t+d+1}{2}\big)_{\frac{d-1}{2}}
\big(\frac{s+t+2d+2}{2}\big)_{\frac{d-1}{2}
}
}{
\big(\frac{s+d+2}{2}\big)_{\frac{d-1}{2}}
\big(\frac{t+d+2}{2}\big)_{\frac{d-1}{2}}
\big(\frac{s+t+2d+1}{2}\big)_{\frac{d-1}{2}}
}
\,.
\end{align}
The relations between partial and full amplitudes admit a further generalization into the following addition-multiplication formula:
\begin{align}
\label{generalizedKLT}
&f^{(d,d')}(s,t)\,
\frac{\sin(\pi s)\sin(\pi t)}{\sin(\pi u)}
A_p^{(d)}(s,t)A_p^{(d')}(s,t)
= 
\\ \nonumber
&
\begin{cases}
A^{(d+d')}(2s,2t) &\text{ for $d+d'$ even,}
\\[8pt]
-\cot(\pi s)\cot(\pi t) A^{(d+d')}(2s,2t) &\text{ for $d+d'$ odd,}
\end{cases}
\end{align}
where the function $f^{(d,d')}(s,t)$ is given by
\begin{align}
f^{(d,d')}(s,t)=\frac{2}{2^{\frac{d+d'}{2}}}h(s)\,h(t)\,H(s+t)\,,
\end{align}
with the functions $h$ and $H$ defined as
\begin{align}
& h(x)=\frac{
\Gamma(\frac{x+2d}{2})
\Gamma(\frac{x+2d'}{2})
\Gamma(\frac{x+d+2}{2})
\Gamma(\frac{x+d'+2}{2})
}{
\prod\limits_{\scriptscriptstyle w\in\{0,1\}}
\begin{matrix}\\[-9pt] \Gamma(\frac{x+d+d'+w}{2})\end{matrix}
\prod\limits_{\scriptscriptstyle w\in\{2,4\}}
\begin{matrix}\\[-9pt] \Gamma(\frac{2x+d+d'+w}{4})\end{matrix}
}\,,
\nonumber \\[-14pt] \\[-4pt]  \nonumber
& H(x)=
\frac{
\prod\limits_{\scriptscriptstyle w\in\{1,2\}}
\begin{matrix}\\[-9pt] \Gamma(\frac{x+d+d'+w}{2})\end{matrix}
\prod\limits_{\scriptscriptstyle w\in\{0,2\}}
\begin{matrix}\\[-9pt]\Gamma(\frac{2x+3(d+d')+w}{4})\end{matrix}
}{
\Gamma(\frac{x+2d+2}{2})
\Gamma(\frac{x+2d'+2}{2})
\Gamma(\frac{x+3d}{2})
\Gamma(\frac{x+3d'}{2})
}
\,.
\end{align}
It can be readily verified that for $d$ and $d'$ odd, $h(x)$ and $H(x)$ are rational functions.

There also exists a different uplift of KLT that expresses the full amplitude as a finite weighted sum over products of two Veneziano partial amplitudes, valid only for even $d$:
\begin{align}
&A^{(d)}(s,t)=\frac{\sin(\pi s)\sin(\pi t)}{\sin\big(\pi(s+t)\big)}
\sum_{n=0}^{d-2}
\binom{d-2}{n}(-1)^{\frac{d+2n}{2}}
\nonumber \\[-14pt] 
\\[-2pt] \nonumber
&
\frac{A_p^{(1)}(s+d-2-n,t+d-2)A_p^{(1)}(s+n,t+d-2)}
{2^{d-2}\pi^{\frac{1-d}{2}}\Gamma(\frac{d-1}{2})}
\,.
\end{align}
One way to derive this equation from \eqref{intgeneral} is to perform a change of variables for the non-gauge fixed Koba-Nielsen variable $\vec{x}=(x_1,x_2,...,x_d)$ by introducing a radial coordinate $r=\sqrt{x_2^2+...+x_d^2}$. After an integral over angular variables, which gives the surface area of the $(d-2)$-sphere, one is left with a two-dimensional integral that can be factorized by going to complex variables $v_\pm=x\pm i r$, just as in the derivation of the standard KLT relation \cite{bjerrum2011momentum} except now the integral contains a factor of $r^{d-2}=(\frac{v_+-v_-}{2i})^{d-2}$, which can be expanded out using the binomial theorem.
\\

\begin{figure}
\centering
\includegraphics[width=0.97\columnwidth]{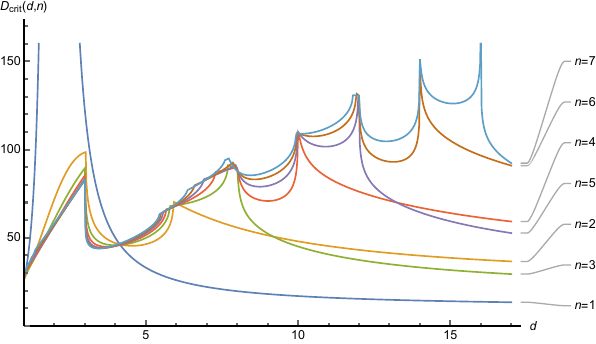}
\caption{\label{fig:Dcrit} 
Critical dimensions $D_\text{crit}(d,n)$ beyond which partial wave coefficients become negative at mass level $n$, plotted for the first six excited levels of the amplitudes $A^{(d)}(s,t)$ obtained from a $d$-dimensional Koba-Nielsen integral. For a given level $n$, inflection points in the curve indicate changes of which spin-$\ell$ coefficients $c_\ell^{(d)}$ become negative for the lowest value of spacetime $D$. For $d=1$ and $d=6$, the lowest value of $D_\text{crit}(n,d)$ equals 26. As $d$ tends to infinity, the lowest value of $D_\text{crit}(n,d)$ asymptotes to 10. 
} 
\end{figure}

\prlsection{Positivity and Critical Dimensions}
Unitarity dictates that the residues of tree-level amplitudes decompose into positively weighted sums of partial waves. In $D$ spacetime dimensions, the spin-$\ell$ partial wave is given by the Gegenbauer polynomial $C_\ell^{(\frac{D-3}{2})}(\cos\theta)$, where $\cos\theta$ is the center-of-mass frame scattering angle. In our case, the partial wave decomposition reads
\begin{align}
-\underset{s=m_n^2}{\text{Res}} A^{(d)}(s,t) = 
\sum_{\ell = 0}^{2n}
c_\ell^{(d)}(n,D)\,C_\ell^{(\frac{D-3}{2})}(\cos\theta)\,.
\label{decomposition}
\end{align}
Since the full amplitudes are symmetric in $t$ and $u$, they are even in $\cos\theta$, and therefore the decomposition \eqref{decomposition} only contains terms with even spin $\ell$. To probe whether the candidate amplitudes $A^{(d)}(s,t)$ might carry any physical significance, let us investigate when, if ever, the positivity conditions for the coefficients $c_\ell^{(d)}(n,D)$ are satisfied. Let us define $D_\text{crit}(d,n)$ as the lowest value of $D$ beyond which, at fixed $d$ and $n$, one or more of the coefficients $c_\ell^{(d)}(n,D)$ become negative. Of course, positivity must be satisfied at all levels $n$, so we can also more properly define 
\begin{align}
D_\text{crit}(d)=\underset{n\in\mathbb{N}}{\text{min}}\,D_\text{crit}(d,n)\,.
\end{align}
Since we cannot check all values of $n$, any value we assign to $D_\text{crit}(d)$ will be conjectural, whereas $D_\text{crit}(d,n)$ can be explicitly computed case by case and perhaps provides an accurate estimate of $D_\text{crit}(d)$. Figure~\ref{fig:Dcrit} displays plots of $D_\text{crit}(d,n)$ for $n$ equal one to seven. For $d$ equal to five and above, the critical dimension appears to be determined by the level $n=1$. The value $D_\text{crit}(d,n)$ with $n=1$ owes to the spin-zero coefficient, 
\begin{align}
c_0^{(d)}(1,D)=
\pi^{d/2}\frac{8+2d(4+5d)-(d-2)^2D}{8(D-1)\Gamma(1+\frac{d}{2})}\,.
\end{align}
Setting this coefficient equal to zero and solving for $D$, we find that
\begin{align}
D_\text{crit}(d,1)
=2\frac{4+4d+5d^2}{(d-2)^2}\,.
\end{align}
Setting $d$ to one gives $D_\text{crit}(1,1)=26$, the critical dimension for the Veneziano amplitude. We recover this same critical dimension by setting $d$ to six: \begin{align}
D_\text{crit}(6,1)=26\,.
\end{align}
To conceive of world volumes propagating inside a larger spacetime, we should perhaps require $d<D$. However, mathematically, there is no obstruction to considering sigma models with a lower-dimensional target space than the world volume, and sometimes dualities play strange tricks on dimensionalities. In any event, taking the large $d$ limit gives
\begin{align}
\lim_{d\rightarrow \infty} D_\text{crit}(d,1) = 10\,.
\end{align}
For $d>2$, the $n=1$ mass level which produces this limiting value is tachyonic. If a consistent procedure exists for excising tachyons from the amplitudes, it will raise the lower bound on the critical dimension. 

It is difficult to make sense of the equation \eqref{intgeneral} for negative $d$, but equation \eqref{Ad} remains meaningful when $d$ is assigned a negative value. However, at negative $d$, the partial wave coefficients become sign-indefinite for any $D$, so these cases are unlikely to be of physical interest. (The same occurs for $d=0$ except in $D=2$, where at least the first many partial wave coefficients all have the same sign.)

While empirical positivity checks of the first many poles do not prove unitarity, a method for rigorously establishing positivity for some ranges of $D$ has been developed in Ref.~\cite{arkani2022unitarity}, through the use of contour integral representations of partial wave coefficients. Their methodology can be straightforwardly carried over and applied to the $d=3$ partial amplitude,
\begin{align}
\mathcal{A}^{(3)}_p(s,t) = 
\frac{2\pi}{(s+4)(t+4)}-2\pi\frac{\Gamma(-s-4)\Gamma(-t-4)}{\Gamma(-s-t-7)}\,.
\end{align}
 The non-zero partial wave coefficients $c_\ell(n,D)$, which have $\ell+n$ even, satisfy the formula
\begin{align}
&c_\ell(n,D)\hspace{-0.5mm}=\hspace{-0.8mm}
\frac{2\,
\Gamma(D-3)
\Gamma(\frac{D-2}{2})
\Gamma(\frac{D+2\ell-2}{2})
\big[\,\Gamma(\frac{D\hspace{-0.5mm}-\hspace{-0.5mm}3}{2})\big]^2}{4^{3-D}(n\hspace{-0.5mm}+\hspace{-0.5mm}1)(n\hspace{-0.5mm}+\hspace{-0.5mm}9)^{D+\ell-3}(2\ell\hspace{-0.5mm}+\hspace{-0.5mm}D\hspace{-0.5mm}-\hspace{-0.5mm}3)^{-1}}
 \nonumber\\[-13pt]
\label{contourFormula}
\\[-3pt] \nonumber
&
\oint \hspace{-0.8mm}dx\hspace{-0.8mm}
\oint \hspace{-0.8mm}dy \frac{(1-x)^{-5}(1-y)^{-5}
\Big(\frac{\frac{1}{\log(1-x)}-\frac{1}{\log(1-y)}}{x-y}\Big)^{\ell}
}{(2\pi i)^2\big(\log(1\hspace{-0.5mm}-\hspace{-0.5mm}x)\log(1\hspace{-0.5mm}-\hspace{-0.5mm}y)\big)^{\frac{D-2}{2}}
\hspace{-1mm}(x\hspace{-0.5mm}-\hspace{-0.5mm}y)^{n+1-\ell}
}
\,,
\end{align}
where the contours wind around the origin. From the fact, proven in \cite{arkani2022unitarity}, that the function $(1-x)^{-1}(-\log(1-x))^\alpha$ has a positive expansion in Gegenbauer polynomials for $\alpha \geq -2$, it follows from \eqref{contourFormula} that the polynomial residues of $A^{(3)}_p(s,t)$ satisfy positivity for $D\leq 22$.
\\

\prlsection{Conclusion}
The preceding sections have uncovered new properties of dimensionally extended string amplitudes and elucidated why we consider them to be a compelling subject matter for physicists to study. The integration formula \eqref{intgeneral} hints at a potential interpretation of $A^{(d)}(s,t)$ as describing the scattering of extended objects through the integration of world-(hyper)volumes ($d$ even, $(d-1)$-dimensional objects) or boundary surfaces thereof ($d$ odd, $d$-dimensional objects). (The amplitudes $A^{(d)}(s,t)$ with $d>2$ could perhaps be termed ``hyperamplitudes"). 

The issue of ghosts is a critical obstacle that obstructs ambitious attempts to leverage such higher-dimensional amplitudes to develop a broadened understanding of tachyon condensation and uplift string field theory to membrane field theory. The infinite-dimensional conformal symmetry for $d\in \{1,2\}$ plays an essential role in excising ghosts from string theory. Green and Thorne \cite{green1991continuing} have argued that the finite-dimensional conformal symmetry for other values of $d$ does not suffice to accomplish this feat, based on an analysis of the residue at the first excited mass level for the most natural six-point generalization of equation \eqref{intgeneral}. However, after many rigorous checks, the four-point amplitude continues to stand out as a healthy physical amplitude to all appearances (when allowing for tachyons) --- and this provides a measure of encouragement to explore potential remedies to the concern raised by Ref.~\cite{green1991continuing}. 

One avenue for trying to incorporate the extended amplitudes into a genuine physical theory free of pathologies would be to search for analogous fermionic and supersymmetric amplitudes along with an enlarged version of the GSO projection \cite{gliozzi1977supersymmetry}. But this line of inquiry is faced with another obstruction, namely the instability of the supermembrane observed by de Wit, Luscher, and Nicolai \cite{deWit:1988xki}. Under a standard kinetic term, M2 branes are susceptible to deformations extending the world-volume into elongated tendrils due to the absence of a penalty associated with length, leading to a contrasting behavior with strings. In Ref.~\cite{gubser2019non} a speculative solution to this challenge was put forward, which set aside the local kinetic term and proposed a non-local but covariant sigma model on general manifolds $M$, grounded in the concept of geodesic arc length $d\big(\phi(x),\phi(y))$ between two values of a field $\phi: M\rightarrow \mathbb{R}^d$. Ref.~\cite{gubser2019non}, however, focused on a bosonic action,
\begin{align}
\label{nonlocalSigma}
S = \frac{-\mu^{d-s}\Gamma(\frac{d+s}{2})}{4\pi^{\frac{d}{2}+s}\hat{\gamma}\Gamma(-\frac{s}{2})} 
\int \frac{d^d\vec{x}\, d^d\vec{y}}{|\vec{x}-\vec{y}|^{d+s}}
d\big(\phi(\vec{x}),\phi(\vec{y})\big)^2\,,
\end{align}
where $\mu$ is an auxiliary length scale, $\hat{\gamma}$ a loop-counting parameter, and $s$ a positive-valued tunable parameter. Taking the $s\rightarrow 2$ limit gives the Polyakov action \cite{deser1976complete,brink1976locally,polyakov1981quantum}, from which the Veneziano amplitude may be derived,
\begin{align}
\lim_{s\rightarrow 2}S = \frac{\mu^{d-2}}{8\pi^2\hat{\gamma}}
\int_{\mathbb{R}^d} d^d\vec{x}\,g_{ab}(\phi) \partial_\mu \phi^a \partial^\mu \phi^b\,.
\end{align}
Establishing a formula like \eqref{intgeneral} for membrane scattering with standard exponential-type vertex operators requires a logarithmic propagator on the world volume. In general, the attainment of a logarithmic correlator for $d>2$ necessitates a higher-derivative kinetic term, and for odd $d$, the requisite kinetic term is non-local, as in \eqref{nonlocalSigma}. Taking the $s\rightarrow d$ limit of this action produces the needed logarithmic world volume propagator.

A comparable situation in physics, to which a parallel may be drawn with the dimensionally extended sigma model, occurs in the case of Liouville theory, which supports a dimensional uplift to non-unitary theories with higher-derivative and non-local kinetic terms \cite{levy2018liouville,kislev2022odd}. The simple $d$-dependence in \eqref{Ad} aligns closely with the generalizations of the DOZZ formula for the Liouville three-point function. 

An alternative pathway to shedding light on the physical significance of the extended amplitudes, which is complementary to top-down explorations founded on specific theories, is offered by the bottom-up approach of the numerical and analytic bootstrap. More broadly, it would be desirable to determine the locations of prospective amplitudes constructed from the higher-dimensional Koba-Nielsen measure within the $S$-matrix landscape that is being charted out by ongoing bootstrap studies, as in Refs.~\cite{cheung2023stringy,haring2023stringy,eckner2024regge,cheung2024bootstrap,berman2024corners}. The formulas for the corrected partial amplitudes given in \eqref{AandA},  \eqref{Rformula}, and \eqref{aandbformula}, and also in equation \eqref{partialMaster} in \hyperref[appendix:C]{Appendix C}, reveal that there are whole new classes of functions to consider when attempting to build amplitudes from the ground up, namely sums of rational functions and gamma function ratios that separately have non-polynomial residues but that when taken together no longer suffer from this issue. 

In short, we hope to have shown that revisiting early dual model attempts to extend worldsheet S-matrices to higher dimensions, from a more modern point of view, offers engaging prospects to search for new physics.
\\

\prlsection{\hspace{23mm}Acknowledgement}

\noindent
We are grateful to Jacob Bourjaily, Changha Choi, Poul Henrik Damgaard, Nick Geiser, Yaron Oz, Fedor Popov, Piljin Yi, and Wayne Zhao for illuminating discussions that helped improve this work. C.~B.~J. thanks the Niels Bohr International Academy for its hospitality during part of this work. N.~E.~J.~B.-B. acknowledges partial support from DFF grant 1026-00077B and the Carlsberg Foundation. The work of C.~B.~J. is supported by the Korea Institute for Advanced Study (KIAS) Grant PG095901.

\bibliography{References}

\setcounter{equation}{0}
\setcounter{figure}{0}
\setcounter{table}{0}
\makeatletter
\renewcommand{\theequation}{A\arabic{equation}}
\renewcommand{\thefigure}{A\arabic{figure}}

\onecolumngrid

\appendix

\section{Appendix A}
\label{appendix:A}
This appendix describes the steps that lead us to identify \eqref{intgeneral} with \eqref{Ad}. We first perform a gauge-fixing of \eqref{intgeneral}, choosing $\vec{x}_a^{\,0}=0$, $\vec{x}_b^{\,0}=\hat{e}$, and $\vec{x}_c^{\,0}=\infty$, where $\hat{e}$ is any unit vector in $\mathbb{R}^d$:
\begin{align}
\label{A1}
& A^{(d)}(s,t) = \int_{\mathbb{R}^d} d^d\vec{x}\,|\vec{x}|^{-2d-s} |\hat{e}-\vec{x}|^{-2d-t}\,.
\end{align}
We will evaluate this integral using manipulations that are a simple variation of the standard way of relating the beta function to a ratio of gamma functions. Recall the analytically extended Fourier transform of a power function $|\vec{x}|^{-\alpha}$:
\begin{align}
\label{Fourier}
\int_{\mathbb{R}^d}d^d\vec{x}\, |\vec{x}|^{-\alpha} e^{2\pi i \vec{\omega}\cdot \vec{x}}
=|\vec{\omega}|^{\alpha-d}\,
\pi^{\alpha-\frac{d}{2}}\,
\frac{\Gamma(\frac{d-\alpha}{2})}{\Gamma(\frac{\alpha}{2})}\,.
\end{align}
We now define a special kind of gamma function and use \eqref{Fourier} to evaluate it:
\begin{align}
\label{gammad}
\Gamma^{(d)}(s)
=\int_{\mathbb{R}^d} d^d\vec{x}\,|\vec{x}|^{-2d-s} e^{2\pi i \hat{e}\cdot \vec{x}}
=\pi^{\frac{3d}{2}+s}\,
\frac{\Gamma(\frac{-s-d}{2})}{\Gamma(\frac{2d+s}{2})}\,.
\end{align}
Taking the product of two gamma functions and formally combining the integrand gives
\begin{align}
\Gamma^{(d)}(s)\,\Gamma^{(d)}(t)
=
\int_{\mathbb{R}^d} d\vec{x}\,d\vec{y}\,|\vec{x}|^{-2d-t}\,|\vec{y}|^{-2d-s} e^{2\pi i\,\hat{e}\cdot (\vec{x}+\vec{y})}
=
\int_{\mathbb{R}^d} d\vec{x}\,d\vec{z}\,|\vec{x}|^{-2d-t}\,|\vec{z}-\vec{x}|^{-2d-s} e^{2\pi i\,\hat{e}\cdot \vec{z}}\,,
\end{align}
where $\vec{z}=\vec{x}+\vec{y}$. The $\vec{x}$-integral is rotationally invariant and so can only depend on the magnitude of $\vec{z}$, and by dimensional analysis, the $\vec{x}$-integral scales as $|\vec{z}|^{-3d-s-t}$. Changing variables from $\vec{x}$ to $\vec{v} = \vec{x}/|\vec{z}|$, the two integrals factorize, with the $\vec{z}$ integral evaluating to a gamma function:
\begin{align}
\label{A5}
\Gamma^{(d)}(s)\,\Gamma^{(d)}(t)=
\Gamma^{(d)}(s+t+d)\int_{\mathbb{R}^d} d\vec{v}\,|\vec{v}|^{-2d-t}\,|\hat{e}-\vec{v}|^{-2d-s} \,.
\end{align}
By \eqref{A1} and \eqref{A5}, we conclude that
\begin{align}
A^{(d)}(s,t) 
= \frac{\Gamma^{(d)}(s)\,\Gamma^{(d)}(t)}{\Gamma^{(d)}(s+t+d)}\,,
\end{align}
which through the use of \eqref{gammad} reproduces \eqref{Ad} as advertised.\\[10pt]

\setcounter{equation}{0}
\setcounter{figure}{0}
\setcounter{table}{0}
\makeatletter
\renewcommand{\theequation}{B\arabic{equation}}
\renewcommand{\thefigure}{B\arabic{figure}}

\section{Appendix B}
\label{appendix:B}

It is a fact, first observed by Witten and Freund \cite{freund1987adelic}, that the full Veneziano amplitude is expressible in terms of Riemann zeta functions. Similarly, the four-gluon amplitude of type-I string theory can be re-expressed in terms of ratios of the Dirichlet $L$-function for the field of Gaussian rationals \cite{ruelle1989adelic}. One generalization of the Riemann zeta function is the hypercubic Epstein zeta function, defined for $\text{Re}[s]>d$ by
\begin{align}
\zeta^{(d)}(s)=
\frac{1}{2}
\sideset{}{'}\sum_{(n_1,n_2,...,n_d)\in \mathbb{Z}^d}\frac{1}{\big(n_1^2+n_2^2+...+n_d^2\big)^{s/2}}\,,
\end{align}
where the prime indicates that the point at the origin is excluded from the sum. For Re$[s]<d$ the function is defined via analytic continuation. Like the Riemann zeta function, the Epstein zeta function satisfies a functional equation:
\begin{align}
\label{EpsteinFunctional}
\pi^{-s/2}\,\Gamma\big(\frac{s}{2}\big)\,
\zeta^{(d)}(s)
=\pi^{\frac{s-d}{2}}\,
\Gamma\big(\frac{d-s}{2}\big)\,
\zeta^{(d)}(d-s)\,.
\end{align}
From \eqref{EpsteinFunctional}, it straightforwardly follows that the amplitudes in \eqref{Ad} can be re-expressed in terms of Epstein zeta functions:
\begin{align}
\hspace{-1.9mm}A^{(d)}(s,t) = \frac{\zeta^{(d)}(2d+s)\,\zeta^{(d)}(2d+t)\,\zeta^{(d)}(2d+u)}{\zeta^{(d)}(-d-s)\,\zeta^{(d)}(-d-t)\,\zeta^{(d)}(-d-u)}\,.
\end{align}
\\[10pt]

\setcounter{equation}{0}
\setcounter{figure}{0}
\setcounter{table}{0}
\makeatletter
\renewcommand{\theequation}{C\arabic{equation}}
\renewcommand{\thefigure}{C\arabic{figure}}

\section{Appendix C}
\label{appendix:C}
In this appendix we describe how the corrected odd-$d$ partial amplitudes can be obtained through a partitioning of the $\mathbb{R}^d$ integration domain in equation \eqref{A1}. The form \eqref{A1} of the amplitude has been obtained from \eqref{intgeneral} by using the conformal symmetry to perform a choice of gauge-fixing. But even in gauge-fixed form, there remains a residual symmetry. In particular, any set of conformal transformations that permutes the gauge-fixing values $0$, $\hat{e}$, and $\infty$ will leave the gauge-fixed amplitude \eqref{A1} invariant. Let us consider a specific example of such a transformation. For the sake of explicitness, we will take the unit vector $\hat{e}$ to be the unit vector in the abscissa direction, $\hat{e}_1=(1,0,0,...,0)$.

By performing a special conformal transformation followed by a translation,
\begin{align}
\vec{x}\rightarrow
\vec{x}\,'=\frac{\vec{x}-\vec{b}\,x^2}{1-2\,\vec{b}\cdot\vec{x}+b^2\,x^2}\,,
\hspace{20mm}
\vec{x}\,'\rightarrow \vec{x}\,''=\vec{x}\,'+\vec{a}\,,
\end{align}
where we choose for both $\vec{a}$ and $\vec{b}$ to be the abscissa unit vector, $\vec{a}=\vec{b}=\hat{e}_1$, we arrive at the combined transformation
\begin{align}
\label{transformation}
\vec{x}=
\left(
\begin{matrix}
x_1
\\
x_2
\\
x_3
\\
\vdots
\\
x_d
\end{matrix}\right)
\hspace{10mm}
\rightarrow 
\hspace{10mm}
\vec{x}\,''=
\frac{1}{1-2x_1+x^2}
\left(
\begin{matrix}
1-x_1
\\
x_2
\\
x_3
\\
\vdots
\\
x_d
\end{matrix}\right)\,.
\end{align}
This transformation is of the above-mentioned type, as it cyclically permutes the origin, unit abscissa, and infinity: 
\begin{align}
\text{
\begin{tikzpicture}
\node at (0,0.7) {$0$};
\node at (0.606218,-0.35) {$\hat{e}_1$};
\node at (-0.606218,-0.35) {$\infty$};
\draw [->] (0+0.13,0.5*0.7+0.11)  -- (0.5*0.606218+0.13,-0.5*0.35+0.11);
\draw [->] (0.5*0.606218,-0.5*0.35-0.18) -- (-0.5*0.606218,-0.5*0.35-0.18);
\draw [->] (-0.5*0.606218-0.13,-0.5*0.35+0.11) -- (0-0.13,0.5*0.7+0.11) ;
\end{tikzpicture}
}\begin{matrix}\\[-40pt].\end{matrix}
\nonumber
\\[-35pt]
\\ \nonumber
\end{align}
The transformation \eqref{transformation} induces a partitioning of the integration domain $\mathbb{R}^d$ into three fundamental domains $(S)$, $(T)$, and $(U)$. The transformation acts locally on each of these regions, in the sense that nearby points are mapped to nearby points, but acts non-locally across the regions. The precise locations of the regions are as follows:
\begin{align}
&(S) = \big\{x_1>\frac{1}{2}\big\} \,\setminus\, \big\{x_1^2+x_2^2+...+x_d^2< 1\big\}\,,  \nonumber \\
&(T) =  \big\{x_1<\frac{1}{2}\big\} \,\setminus\, \big\{(1-x_1)^2+x_2^2+...+x_d^2< 1\big\}\,,  \\[4pt] \nonumber
&(U) = \big\{x_1^2+x_2^2+...+x_d^2< 1\big\}\,\cap\,\big\{(1-x_1)^2+x_2^2+...+x_d^2< 1\big\}\,.
\end{align}
\begin{figure}
\centering
\begin{align*}
\begin{matrix}
\text{\scalebox{0.8}{
\begin{tikzpicture}
\draw[thick] (0,0) ellipse (0.7cm and 0.7cm);
\filldraw[white] (0,0.7) ellipse (0.25cm and 0.25cm);
\filldraw[white] (0.606218,-0.35) ellipse (0.25cm and 0.25cm);
\filldraw[white] (-0.606218,-0.35) ellipse (0.25cm and 0.25cm);
\filldraw[col6] (0,0.7) ellipse (0.1cm and 0.1cm);
\filldraw[col4] (0.606218,-0.35) ellipse (0.1cm and 0.1cm);
\filldraw[col5] (-0.606218,-0.35) ellipse (0.1cm and 0.1cm);
\draw [very thick,->] (0.64, 0.27) -- (0.64+0.0015, 0.27-0.003);
\draw [very thick,->] (0, -0.7) -- (-0.001, -0.7);
\draw [very thick,->] (-0.56, 0.41) -- (-0.56+0.0015, 0.41+0.002);
\end{tikzpicture}
}}
\end{matrix}
\hspace{12mm}
\begin{matrix}
\text{
\includegraphics[width=0.45\columnwidth]{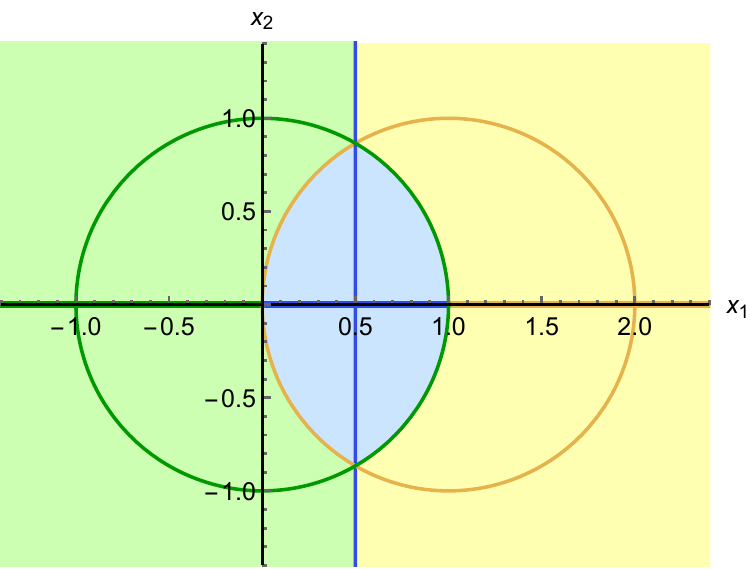}}
\end{matrix}
\hspace{10mm}
\begin{matrix}
\text{\scalebox{0.8}{
\begin{tikzpicture}
\draw[thick] (0,0) ellipse (0.7cm and 0.7cm);
\filldraw[white] (0,0.7) ellipse (0.25cm and 0.25cm);
\filldraw[white] (0.606218,-0.35) ellipse (0.25cm and 0.25cm);
\filldraw[white] (-0.606218,-0.35) ellipse (0.25cm and 0.25cm);
\filldraw[col3] (0,0.7) ellipse (0.1cm and 0.1cm);
\filldraw[col2] (0.606218,-0.35) ellipse (0.1cm and 0.1cm);
\filldraw[col1] (-0.606218,-0.35) ellipse (0.1cm and 0.1cm);
\draw [very thick,->] (0.64, 0.27) -- (0.64+0.0015, 0.27-0.003);
\draw [very thick,->] (0, -0.7) -- (-0.001, -0.7);
\draw [very thick,->] (-0.56, 0.41) -- (-0.56+0.0015, 0.41+0.002);
\end{tikzpicture}
}}
\end{matrix}
\end{align*}
\caption{Plot of the fundamental domains of the transformation \eqref{transformation}. Blue gets mapped to yellow, yellow to green, and green to blue. In dimensions $d$ higher than two, the domains are rotationally symmetric around the $x_1$ axis.
\label{fig:Regions} 
} 
\end{figure} 
In Figure~\ref{fig:Regions} we depict the sections of these domains that lie in the $(x_1,x_2)$-plane.  Breaking the $\mathbb{R}^d$ integral in \eqref{A1} into integrals over these three sub-domains provides a way of splitting the full amplitude into partial amplitudes. While the full integral does not converge, there are kinematic regimes where the integral over one sub-domain converges. In particular, we can consider a $u$-channel regime where $s,t<-d$. Then the $(U)$ integral converges, and by analytic continuation of this integral, the integrals over the other sub-domains can be obtained according to the following identifications:
\begin{align}
&\int_{(U)}d^d\vec{x}\,|\vec{x}|^{-2d-s} |\hat{e}_1-\vec{x}|^{-2d-t}\equiv \mathcal{A}_p^{(d)}(s,t)\,,
\nonumber \\
&\int_{(T)}d^d\vec{x}\,|\vec{x}|^{-2d-s} |\hat{e}_1-\vec{x}|^{-2d-t} =
\int_{(U)}d^d\vec{x}\,|\vec{x}|^{2d+s+t} |\hat{e}_1-\vec{x}|^{-2d-s} =
\mathcal{A}_p^{(d)}(u,s)\,, \label{channelIntegrals}
\\ \nonumber 
&\int_{(S)}d^d\vec{x}\,|\vec{x}|^{-2d-s} |\hat{e}_1-\vec{x}|^{-2d-t} = 
\int_{(U)}d^d\vec{x}\,|\vec{x}|^{-2d-t} |\hat{e}_1-\vec{x}|^{2d+s+t} =
\mathcal{A}_p^{(d)}(t,u)\,.
\end{align}
Visually, we can depict the way the transformation \eqref{transformation} permutes the three fundamental domains and the partial amplitudes with the following diagrams:
\begin{align}
\text{
\begin{tikzpicture}
\node at (0,0.75) {$(U)$};
\node at (0.65,-0.35) {$(S)$};
\node at (-0.65,-0.35) {$(T)$};
\draw [->] (0+0.13,0.5*0.7+0.11)  -- (0.5*0.606218+0.13,-0.5*0.35+0.11);
\draw [->] (0.5*0.606218,-0.5*0.35-0.18) -- (-0.5*0.606218,-0.5*0.35-0.18);
\draw [->] (-0.5*0.606218-0.13,-0.5*0.35+0.11) -- (0-0.13,0.5*0.7+0.11) ;
\end{tikzpicture}
}\begin{matrix}\\[-40pt],\end{matrix}
\hspace{20mm}
\text{
\begin{tikzpicture}
\node at (0,0.8) {$\mathcal{A}_p^{(d)}(s,t)$};
\node at (1.2,-0.35) {$\mathcal{A}_p^{(d)}(t,u)$};
\node at (-1.2,-0.35) {$\mathcal{A}_p^{(d)}(u,s)$};
\draw [->] (0+0.13,0.5*0.7+0.11)  -- (0.5*0.606218+0.13,-0.5*0.35+0.11);
\draw [->] (0.5*0.606218,-0.5*0.35-0.18) -- (-0.5*0.606218,-0.5*0.35-0.18);
\draw [->] (-0.5*0.606218-0.13,-0.5*0.35+0.11) -- (0-0.13,0.5*0.7+0.11) ;
\end{tikzpicture}
}\begin{matrix}\\[-40pt].\end{matrix}
\nonumber
\\[-35pt]
\\ \nonumber
\end{align}
In consequence, we arrive at a decomposition of the full amplitude into partial amplitudes,
\begin{align}
A^{(d)}(s,t)=
\bigg(\int_{(U)}+\int_{(T)}+\int_{(S)}\bigg)\,
d^d\vec{x}\,|\vec{x}|^{-2d-s} |\hat{e}_1-\vec{x}|^{-2d-t}
=
\mathcal{A}_p^{(d)}(s,t)+
\mathcal{A}_p^{(d)}(s,u)+
\mathcal{A}_p^{(d)}(t,u)\,.
\end{align}
Within the convergent region in the $u$-channel, the partial amplitude is given by the integral
\begin{align}
\mathcal{A}^{(d)}_p(s,t)=2^{d-1}
\bigg(&
\int_0^{1/2} dx_1 
\int_0^{\sqrt{x_1(2-x_1)}} dx_2
\int_0^{\sqrt{x_1(2-x_1)-x_2^2}} dx_3
...
\int_0^{\sqrt{x_1(2-x_1)-x_2^2-...-x_{d-1}^2}} dx_d
\nonumber \\
&+
\int_0^{1/2} dx_1 
\int_0^{\sqrt{1-x_1^2}} dx_2
\int_0^{\sqrt{1-x_1^2-x_2^2}} dx_3
...
\int_0^{\sqrt{1-x_1^2-x_2^2-...-x_{d-1}^2}} dx_d
\bigg)
\\ \nonumber 
&\hspace{-2.5mm}
\Big(x_1^2+x_2^2+...+x_d^2\Big)^{-d-s/2}
\Big((1-x_1)^2+x_2^2+...+x_d^2\Big)^{-d-t/2}\,.
\end{align}
By switching to cylindrical coordinates and carrying out the angular integration, we obtain a simpler integration formula for the partial amplitude,
\begin{align}
\mathcal{A}^{(d)}_p(s,t)=
\frac{\pi^{\frac{d-1}{2}}}{\Gamma(\frac{d-1}{2})}
\int_0^{1/2} dx 
\int_0^{x(2-x)} dv\,v^{\frac{d-3}{2}}\,
\Big(x^2+v\Big)^{-d-s/2}
\Big((1-x)^2+v\Big)^{-d-t/2}
+(s\leftrightarrow t)\,.
\end{align}
This formula makes sense only for $d\neq 1$, but the equations \eqref{channelIntegrals} apply also for $d=1$. By carrying out the partial amplitude integral for odd values of $d$, we arrive at the master formula for the corrected partial amplitudes:
\begin{align}
\label{partialMaster}
&\mathcal{A}_p^{(d)}(s,t)=-(16\pi)^{\frac{d}{2}}
\big(\frac{s+d+2}{2}\big)_{\frac{d-3}{2}}
\big(\frac{t+d+2}{2}\big)_{\frac{d-3}{2}}
\big(\frac{s+t+2d+3}{2}\big)_{\frac{d-3}{2}}
\frac{\Gamma(-s-2d+2)\Gamma(-t-2d+2)}{\sqrt{1024\pi}\,\Gamma(-s-t-2d-1)}
\\ \nonumber
&
-\frac{(-\pi)^{\frac{d-1}{2}}}{2^{d-4}\Gamma(\frac{d-1}{2})}
\bigg(
\sum_{M=0}^{\frac{d-3}{2}}
\frac{4^M\left(\frac{3-d}{2}\right)_M}{M!(s+4+2M)(t+4+2M)}-
4
\sum_{M=0}^{\frac{d-5}{2}}
\frac{(3-d)_M}{M!\left(\frac{4-d}{2}\right)_M(s+2d-2-2M)(t+2d-2-2M)}\times
\\ \nonumber
&\hspace{82.5mm} 
\sum_{n=0}^{d-4-2M}
\frac{(d-n-3-2M)^2\left(\frac{4-d+2n}{2}\right)_M(3+M-d)_n}{n!(s+4+2M+2n)(t+4+2M+2n)}
\bigg)
\,.
\end{align}

\end{document}